\documentclass[a4paper,11pt]{article}
\pdfoutput=1 % if your are submitting a pdflatex (i.e. if you have
\usepackage{jcappub} % for details on the use of the package, please
\usepackage[T1]{fontenc} % if needed
\usepackage{comment}
\usepackage{xspace}
\usepackage{physics}

\usepackage{lineno}

\usepackage{ulem}

\usepackage{booktabs}
\renewcommand{\arraystretch}{1.2}

\usepackage{xcolor}

\def\d{{\rm d}}
\newcommand{\Sem}{S_{\rm em}}
\newcommand{\Smu}{S_{\mu}}
\newcommand{\SWCD}{S_{\rm WCD}}
\newcommand{\SSSD}{S_{\rm SSD}}

\title{\boldmath Identification of ultra-high-energy neutrinos in vertical air showers with surface detectors}

\author[a]{Jaime Alvarez-Mu\~niz}
\author[b,c]{Ruben Concei\c{c}\~ao}
\author[b,c]{Pedro J. Costa}
\author[b,c]{David Dias}
\author[b,c]{M\'ario Pimenta}

\affiliation[a]{Instituto Galego de F\'isica de Altas Enerx\'ias (IGFAE), Universidade de Santiago de Compostela, 15782 Santiago de Compostela, Spain}
\affiliation[b]{Departamento de F\'isica, Instituto Superior T\'{e}cnico (IST),\\ Universidade de Lisboa, Av. Rovisco Pais 1, 1049-001, Lisbon, Portugal}
\affiliation[c]{Laboratório de Instrumentação e Física Experimental de Partículas (LIP),\\ Av. Prof. Gama Pinto, 2, P-1649-003 Lisbon, Portugal}

\emailAdd{pmcosta@lip.pt}

\abstract{
We investigate the potential of surface detector arrays to identify ultra-high-energy neutrinos using vertical air showers with zenith angle $\theta \leq 60^\circ$, a region largely unexplored in current searches. The method exploits the enhanced electromagnetic component at ground level arising from energetic leptons produced in neutrino interactions. 
Using detailed air-shower and detector simulations, we show that observables sensitive to the electromagnetic and muonic content provide genuine discrimination between neutrino-induced and hadron-induced showers. This is demonstrated both with idealized shower quantities and with realistic detector-level signals based on water-Cherenkov and scintillator detectors. The corresponding neutrino identification efficiency is used to derive the effective mass and, primarily, the effective area of the array under background-rejection requirements appropriate to targeted multi-messenger searches.
Although not competitive with inclined-shower searches under the stringent background rejection factors required for diffuse-flux analyses, this approach establishes the feasibility of a vertical detection channel, extending the field of view of surface arrays and providing a complementary tool for multi-messenger observations of transient sources.
}

\begin{document}
\maketitle
\flushbottom

%%%%%%%%%%%%%%%%%%%%%%%%%%%%%%%%%%%%%%%%%%%%%%%%%%%%%%%%%%%%%%%%%%
\section{Introduction}
\label{sec:intro}

Neutrino astronomy provides a unique window into the most extreme phenomena in the Universe. In particular, the detection of ultra-high-energy (UHE) neutrinos, with energies above $\sim 10^{17}\,$eV, would provide crucial information on the long-standing question of the origin of the UHE cosmic rays, on the environments in which hadronic acceleration takes place, and on the interactions undergone by these particles both at their sources and during propagation. UHE neutrinos are expected to arise both from the interaction of cosmic rays with cosmic microwave background photons, leading to the so-called cosmogenic neutrino flux~\cite{Berezinsky1969}, and from interactions taking place inside powerful astrophysical accelerators such as active galactic nuclei and gamma-ray bursts~\cite{Halzen2002}. Combined with observations of photons, cosmic rays, and gravitational waves, they constitute a fundamental component of the multi-messenger approach to the high-energy Universe \cite{Guepin:2022qpl}. In particular, coincident observations of ultra-high-energy neutrinos and gravitational waves have been proposed as a powerful probe of transient sources, such as binary neutron-star mergers \cite{Farrar:2024zsm}. Neutrinos are especially valuable in this context because, being electrically neutral and weakly interacting, they can reach the Earth from dense and distant environments that may be opaque to photons.

Several techniques are being exploited to detect UHE neutrinos \cite{MammenAbraham:2022xoc, Ackermann:2022rqc}. Among them, searches for UHE neutrinos with large surface particle detector arrays rely primarily on very inclined air showers~\cite{Auger_DownGoing, Auger_EarthSkimming, SWGOneutrinoInclined} with zenith angle $\theta\geq 60^\circ$. In this regime, the efficiency of neutrino identification is enhanced. Neutrinos can penetrate deeply into the atmosphere before interacting and develop showers close to the ground with a significant electromagnetic component. In contrast, hadronic cosmic rays interact in the upper layers of the atmosphere and at large zenith angles produce showers that are mainly dominated by the muonic component at ground level.
The Pierre Auger Observatory, for instance, exploits two such channels: the Earth-skimming channel, sensitive mainly to tau neutrinos interacting in the Earth's crust and producing upward-going showers, and the downward-going channel, which targets neutrinos of all flavors interacting in the atmosphere at large zenith angles \cite{Auger_EarthSkimming, Auger_DownGoing}.
In both cases, discrimination against the overwhelming cosmic-ray background relies on the difference between ordinary hadronic showers at large zenith angle, which are predominantly muonic, and deeply initiated neutrino-induced showers, which are electromagnetic-dominated.

In contrast, showers with zenith angles $\theta < 60^\circ$ are not usually considered in neutrino searches.
In this angular range, the smaller atmospheric depth traversed by the shower makes it more difficult to disentangle neutrino-induced events from hadronic ones. 
As a consequence, a large fraction of the field of view of surface detector arrays remains unexplored for neutrino detection, thereby limiting their multi-messenger reach.

Nevertheless, interactions of electron and muon neutrinos in the atmosphere can still produce vertical showers whose particle content at ground differs from that of ordinary cosmic-ray-induced showers. In particular, their electromagnetic-to-muonic ratio can depart from that expected for the hadronic background, thereby providing a possible handle for neutrino identification \cite{SWGOneutrinoInclined}.

Building on this idea, in this work we investigate the discrimination power of observables related to the electromagnetic and muonic components of the shower at ground level, for the search for neutrinos in the zenith angle range $\theta < 60^\circ$. Firstly, we study the discrimination at the level of idealized Monte Carlo quantities and then using detector-level proxies accessible to water-Cherenkov- and scintillator-based surface arrays. We consider both electron- and muon-neutrino induced showers and evaluate the corresponding identification efficiency as a function of the interaction depth in the atmosphere.

 The main result of this work is that the electromagnetic and muonic content at ground level provides genuine separation power between vertical neutrino-induced showers and the cosmic-ray background. This demonstrates the feasibility of using air showers with $\theta < 60^\circ$ as a neutrino-search channel in surface detector arrays, thereby expanding the sky coverage of current and future particle detector arrays where these components, or suitable proxies for them, can be experimentally accessed. Building on these results, we estimate the corresponding effective mass and effective area and derive the associated flux sensitivity for a ground-based array. This study is framed as a proof of principle. Its goal is not to compete with the existing, more efficient searches for neutrinos based on inclined air showers, but rather to explore the complementary possibility of identifying more vertically developing neutrino-induced air showers in coincidence with multimessenger observations, thereby extending the multimessenger capabilities of surface detector arrays.

This paper is organized as follows. In Section~\ref{sec:simulation}, we describe the simulation framework used in this work. In Section~\ref{subsec:ideal}, we present a proof-of-principle analysis based on simulated electromagnetic and muonic signals at ground level. The analysis is then extended in Section~\ref{sec:exp_sd} to realistic detector-level observables using the water-Cherenkov and scintillator surface detector responses. In Section~\ref{sec:sensitivity}, we obtain the effective mass and effective area for the search for vertical neutrinos and derive the corresponding sensitivity to UHE neutrinos.

In Section~\ref{sec:neutrals}, we discuss the role of gamma-induced showers as a background for neutral-particle searches in the vertical channel.
Finally, in Section~\ref{sec:conclusions}, we summarize our results and discuss their implications for future neutrino searches with surface detector arrays.

%%%%%%%%%%%%%%%%%%%%%%%%%%%%%%%%%%%%%%%%%%%%%%%%%%%%%%%%%%%%%%%%%%
\section{Simulation framework}
\label{sec:simulation}

To investigate the potential of vertical air showers\footnote{In this work, the term \textit{vertical} refers to showers with zenith angles between $0^\circ$ and $60^\circ$.} for UHE neutrino detection, a dedicated simulation framework was developed to model both the development of extensive air showers and the response of a ground-based detector array. In order to establish a proof of principle and isolate the impact of the electromagnetic and muonic shower components at ground level, the simulations were restricted to a fixed zenith angle and a single primary energy for neutrinos. A full treatment including the energy and angular dependence of the identification efficiency would require significantly larger simulation statistics and is deferred to a future work.

All quantitative results in this work should therefore be read as representative of the energy around $\sim 10^{18}\,\mathrm{eV}$ and zenith angle $\theta\simeq 30^\circ$ adopted in the simulations, rather than as a full instrument response function. Where extrapolations in energy or zenith angle are required for the sensitivity estimate, they are explicitly stated and treated conservatively.

%%%%%%%%%%%%%%%%%%%%%%%%%%%%%%%%%%%%%%%%%%%%%%%%%%%%%%%%%%%%%%%%%%
\subsection{Air-shower simulations}

Extensive air showers initiated by neutrinos and hadronic cosmic rays were simulated using the \textsc{CORSIKA} package~\cite{CORSIKA}. The EPOS-LHC model~\cite{2015_Pierog_eposlhc} was used for high-energy hadronic interactions, while low-energy interactions were handled by \textsc{FLUKA}~\cite{fluka}.

Neutrino as well as photon-induced showers were simulated at energy $\log(E/{\rm eV}) = 18$, corresponding to the energy range where the $1500\,{\rm m}$-spaced surface detector array considered in this work (see section~\ref{sec:det_array}) becomes sensitive to vertical neutrino-induced showers. 
This energy value is also representative of the energy range where the Pierre Auger Observatory, a realistic reference for the detector array considered in this work, is expected to have the highest sensitivity to UHE neutrinos~\cite{Auger_EarthSkimming}.
Conversely, proton primaries, used as the background sample, were simulated with energies sampled uniformly in $\log(E/{\rm eV})$ over the range $\log(E/{\rm eV}) \in [17.3,\,18.3]$, ensuring that the background sample covers the reconstructed-energy window relevant to a $10^{18}\,{\rm eV}$ neutrino search. The lower edge of the simulated interval accounts for the tendency of electromagnetic-dominated showers to be reconstructed with a lower energy than their true primary energy~\cite{PierreAuger:2025jwt}.
All showers were simulated at a fixed zenith angle of $\theta=30^\circ$, representative of vertical events.

Both electron and muon neutrinos were simulated, and their interactions in the atmosphere were modeled using the \textsc{HERWIG} package~\cite{HERWIG}. Due to the very small neutrino interaction cross-sections, the interaction depth must be explicitly sampled. Interaction points were therefore generated uniformly in atmospheric grammage, from the top of the atmosphere down to the detector level, corresponding to a vertical depth $X_{\rm array}\simeq 875.5\,{\rm g\,cm^{-2}}$ and an altitude of $\approx 1400\,{\rm m}$ a.s.l. The atmosphere was split into $18$ bins of $36\,{\rm g\,cm^{-2}}$, and for each bin, approximately $300$ neutrino showers were simulated to ensure adequate statistics.

Only showers that triggered the detector array were retained for the analysis, as these represent the subset of neutrino events that would be detectable in a real experimental scenario.

For each primary configuration (defined by particle type, energy, zenith angle and, for neutrinos, interaction depth), a maximum of 10 independent showers was simulated. Shower generation was stopped as soon as one shower satisfied the array trigger. This procedure accounts for shower-to-shower fluctuations while ensuring that the subsequent identification analysis is performed only on triggered events. If none of the 10 simulated showers triggered the array, the corresponding configuration was considered undetectable and assigned zero trigger efficiency.

The shower core position was fixed to coincide with the central station of the array for all simulated events, due to computational limitations. We note that this fixed geometry does not necessarily provide an optimal detector response, particularly when accounting for station saturation. The resulting trigger efficiencies should therefore be regarded as conservative estimates.

%%%%%%%%%%%%%%%%%%%%%%%%%%%%%%%%%%%%%%%%%%%%%%%%%%%%%%%%%%%%%%%%%%
\subsection{Detector array simulation}
\label{sec:det_array}

The detector response was modeled using a \textsc{Geant4}-based simulation~\cite{Geant4} of a surface detector array composed of water-Cherenkov detectors (WCDs) arranged on a triangular grid with a spacing of $1500\,{\rm m}$, similar to the layout of the main array of the Pierre Auger Observatory~\cite{Auger}. The array spans a circular area with a radius of $9\,{\rm km}$ which, at the simulated zenith angle of $\theta=30^\circ$, is sufficiently large to ensure that the footprints of showers whose cores fall near the centre are fully contained within the array's boundaries. For instance, a $E=10^{18}~{\rm eV}$ proton shower with a zenith angle of $30^\circ$ has a lateral spread of approximately $1.1$--$1.9\,{\rm km}$ at ground level~\cite{Gaisser2016CosmicRays}, which is well within the $9\,{\rm km}$ radius of the array.

\begin{figure}[h!]
    \centering
    \includegraphics[width=0.5\textwidth]{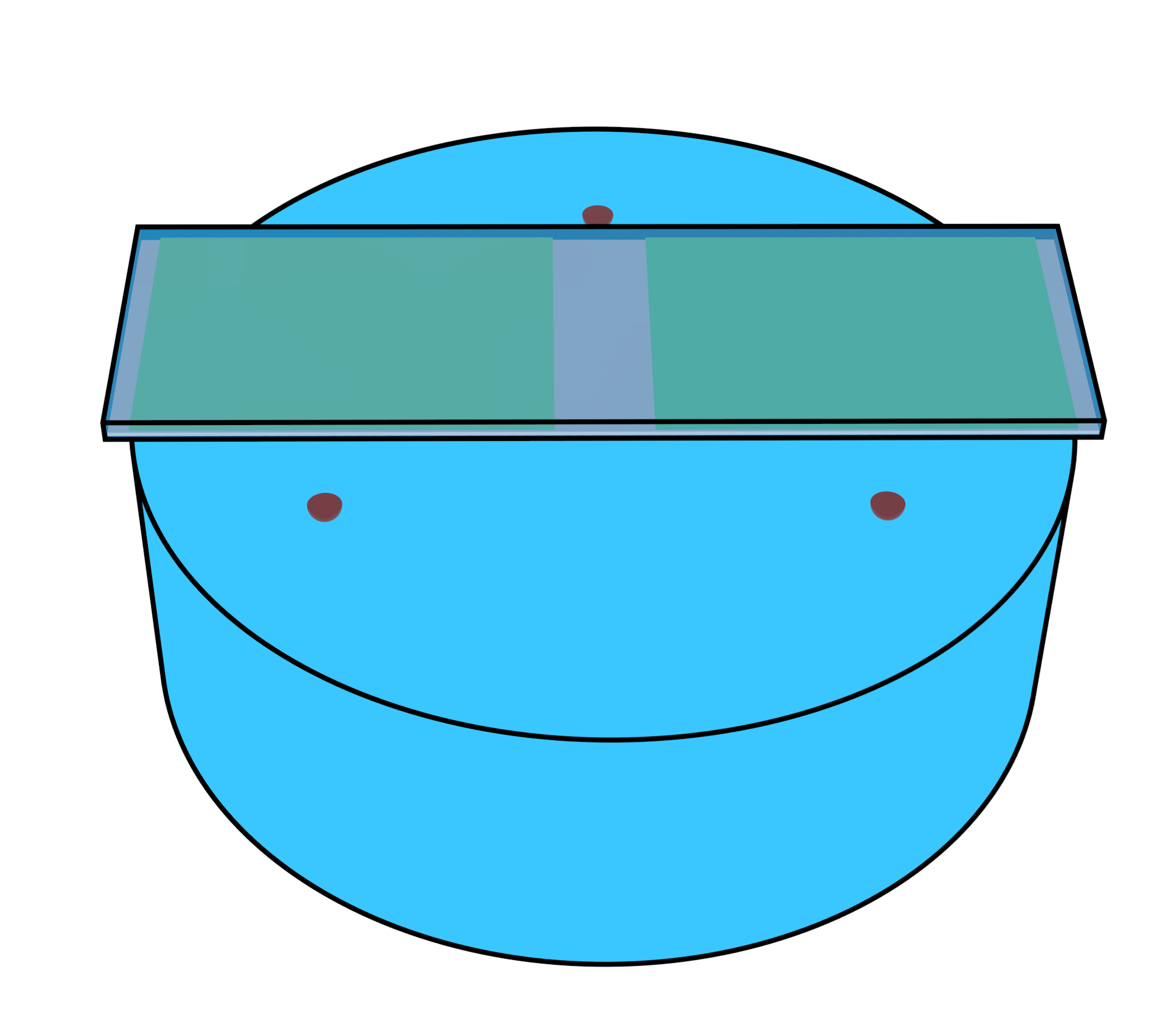}
    \caption{
    Sketch of the simulated detector station, consisting of a water Cherenkov detector (WCD) with three photomultiplier tubes (PMTs) at the top (red dots), and a rectangular scintillator surface detector (SSD) on top of the WCD.}
    \label{fig:WCDStation}
\end{figure}

Each detector station consists of a cylindrical water Cherenkov detector, $1.2\,{\rm m}$ in height and $3.6\,{\rm m}$ in diameter, equipped with three photomultiplier tubes (PMTs). 
A scintillator surface detector (SSD) is placed on top of each WCD, consisting of two scintillator planes with a total active area of $4\,{\rm m^2}$. A sketch of the simulated detector station is shown in Figure~\ref{fig:WCDStation}. The SSDs are primarily sensitive to the electromagnetic component of the shower with a relatively weak dependence on shower particle type, while the WCDs respond to both the electromagnetic and muonic components, possessing a larger relative response to muons. This complementary response is essential for estimating the relative electromagnetic and muonic content of the shower at ground level.

%%%%%%%%%%%%%%%%%%%%%%%%%%%%%%%%%%%%%%%%%%%%%%%%%%%%%%%%%%%%%%%%%%
\subsection{Trigger and shower sampling at ground}

Shower cores were fixed at the center of the array for all simulated events in order to minimize border effects and reduce shower-to-shower fluctuations associated with different core positions. A station-level trigger was implemented by requiring each of the three PMTs in a WCD to record a total signal above a threshold of $1.2\,{\rm VEM}$ (Vertical Equivalent Muon) \cite{Auger, Auger_SD_Aperture_and_Trigger}. An event was considered to trigger the array if at least three stations satisfied this condition. 

Although this trigger threshold is lower than the standard Auger SD station threshold of $1.75\,{\rm VEM}$~\cite{Auger_SD_Aperture_and_Trigger}, it was chosen to ensure that a sufficient number of neutrino-induced showers at $10^{18}\,{\rm eV}$ trigger the array, allowing for a meaningful proof-of-principle analysis. In a realistic experimental scenario, a more sophisticated trigger or a dedicated low-threshold strategy optimized for enhanced sensitivity to the electromagnetic component of the shower would be required to suppress random coincidences.
The results presented here should therefore be interpreted as an upper bound on the achievable identification performance at this energy, pending a full trigger-optimisation study that is deferred to a future work.

In experimental data, the shower signals at ground level are typically extracted by fitting the lateral distribution function (LDF) of the shower. To focus strictly on the intrinsic physics differences between neutrino-induced and hadron-induced showers, and to avoid the additional uncertainties introduced by LDF fitting, a dense ring of virtual detector stations was placed at a radial distance of $1000\,{\rm m}$ from the shower core along the shower plane perpendicular to the shower axis. The spacing between virtual stations was set to $10\,{\rm m}$. This dense sampling allows for a precise determination of the average shower signals at a fixed distance from the core, effectively providing the idealised information that would otherwise be extracted from the LDF fits. The impact of realistic LDF reconstruction is discussed qualitatively in Section~\ref{sec:conclusions}.

%%%%%%%%%%%%%%%%%%%%%%%%%%%%%%%%%%%%%%%%%%%%%%%%%%%%%%%%%%%%%%%%%%
\section{Methodology and Ideal Monte Carlo Discrimination} 
\label{sec:proof_of_principle}

The primary feature used to distinguish neutrino-induced showers from hadronic vertical showers is the relative proportion of electromagnetic and muonic component present at ground level. Hadronic showers are initiated in the upper layers of the atmosphere and consequently exhibit a larger muonic component at ground relative to their electromagnetic component than neutrino-induced showers. This is primarily due to the decay of charged pions and kaons into penetrating muons as the shower develops, while the electromagnetic component is progressively built through neutral pion production and more rapidly attenuated through processes such as pair production, bremsstrahlung, and ionization losses.

In contrast, neutrino-induced showers are initiated through weak interactions anywhere along the atmosphere with the same probability. In the specific case of electron-neutrinos and charged-current interactions, a highly energetic electron is produced, transferring on average roughly $70$--$80\%$ of the primary neutrino energy into an electromagnetic cascade. While the fragmentation of the target nucleon generates an unavoidable hadronic sub-shower that behaves similarly to a standard hadronic cascade, the initial injection of energy into the electromagnetic channel results in a shower with a significantly larger ratio of electromagnetic to muonic signal at the ground compared to standard hadronic showers.

Muon neutrinos are expected to provide a less efficient detection channel relative to electron neutrinos within this approach.
The high-energy muon produced in a charged-current interaction can contribute to the electromagnetic component through radiative processes such as bremsstrahlung, provided its energy remains sufficiently high (typically above $\sim 100\,{\rm TeV}$)~\cite{pdg}.
However, once muon energy losses become dominated by ionisation rather than radiative processes, the muon ceases to contribute significantly to the electromagnetic cascade, and the ground-level electromagnetic-to-muonic ratio approaches that of ordinary hadronic showers. Additionally, the shower initiated by the hadronic system produced in the muon-neutrino interaction with the target nucleus carries on average only $20$--$30\%$ of the primary neutrino energy, and remains predominantly hadronic.

Regardless of the potentially distinctive signatures of neutrinos interacting close to the detector, the resulting showers must still trigger the array. This requires the first interaction to occur sufficiently high in the atmosphere to allow for adequate shower development. The identification of neutrino-induced vertical showers therefore relies on a balance between the interaction depth and the shower characteristics at ground level.

%%%%%%%%%%%%%%%%%%%%%%%%%%%%%%%%%%%%%%%%%%%%%%%%%%%%%%%%%%%%%%%%%%
\subsection{Discrimination in the Ideal Case}
\label{subsec:ideal}

As an initial proof-of-principle analysis corresponding to an ideal scenario, we directly use the simulated average signals of the pure electromagnetic ($\Sem$) and muonic ($\Smu$) components. These values are extracted at $1000\,{\rm m}$ from the shower core along the shower plane, averaged over the detectors at the densely sampled ring of virtual stations described in Section~\ref{sec:simulation}.

Since the simulated proton events follow a continuous energy distribution, all discriminant variables associated with these events are weighted according to the cosmic-ray flux measured by the Pierre Auger Observatory~\cite{Auger_CR_Spectrum_ICRC2021}, $\Phi(E) \propto E^{-\gamma}$, with $\gamma \approx 2.7$ in the energy range of interest. This ensures a realistic representation of the expected background spectrum.

\begin{figure}[htbp]
    \centering
    \includegraphics[width=0.6\textwidth]{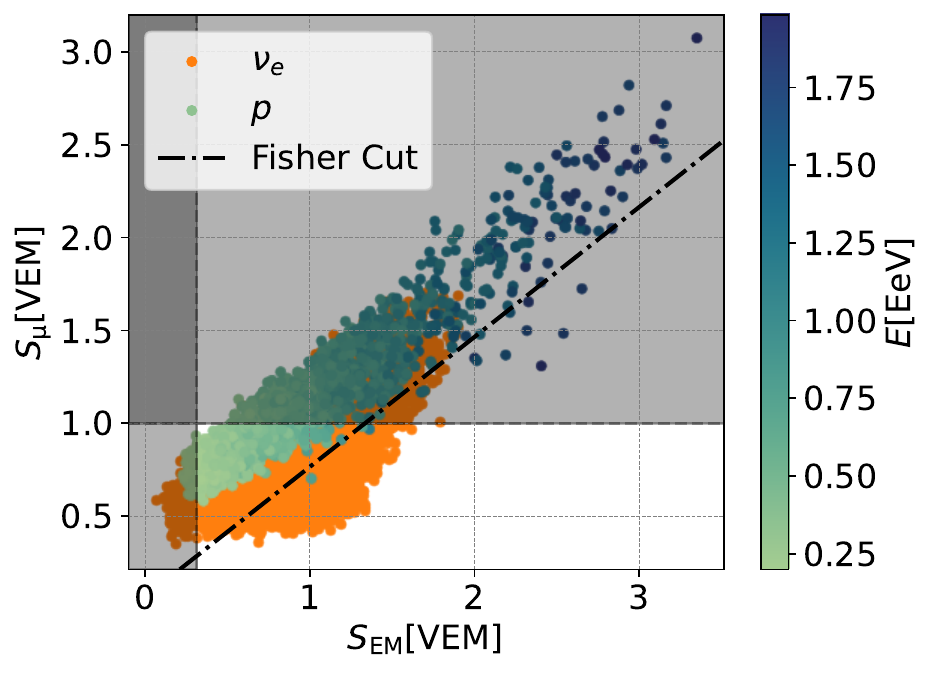}
    \caption{Scatter plot of the average electromagnetic signal $\Sem$ and muonic signal $\Smu$ measured by the dense ring of stations at a distance $r = 1000\,{\rm m}$ from the shower core. Electron-neutrino-induced showers with $E=10^{18}\,\mathrm{eV}$ and $\theta=30^\circ$ are shown as orange markers, while proton-induced showers are color-coded according to the primary energy (color scale on the right). The dashed horizontal and vertical lines indicate the 10th percentile cut in $\Sem$ and the 90th percentile cut in $\Smu$ for proton events, weighted by the cosmic-ray flux. These cuts define the exclusion region for background events, shown as the grey shaded area. The dash-dotted line corresponds to the adjusted Fisher cut for a proton background rejection factor of $5\times10^{4}$.
    }
    \label{fig:Scatter_Sem_Vs_Smu}
\end{figure}

Two initial geometric cuts were applied to the $\Sem$ and $\Smu$ variables to define a region of interest, as illustrated in Figure~\ref{fig:Scatter_Sem_Vs_Smu}. Because hadronic showers are generally more muon-rich, we require the muonic signal $\Smu$ to be below the $90^\mathrm{th}$ percentile of the flux-weighted proton $\Smu$ distribution, thereby rejecting the $10\%$ most muon-rich proton events. Conversely, since neutrino-induced showers are expected to retain a larger electromagnetic component, we require the electromagnetic signal $\Sem$ to exceed the $10^\mathrm{th}$ percentile of the corresponding proton distribution, rejecting the $10\%$ of proton events with the weakest electromagnetic signals.
The combined introduction of these \textit{rectangular cuts} results in the exclusion of around $20\%$ of the proton background. For electron neutrinos, approximately $15\%$ of all signal events are excluded, with the highest and lowest interaction depths simulated being the most affected, losing up to $20\%$ of events initiated at the corresponding heights above ground. 
While simple, these cuts efficiently remove the bulk of the uninteresting proton background, as shown in Figure~\ref{fig:Scatter_Sem_Vs_Smu}, and they allow us to retain a subset of events most similar to neutrino-induced showers in terms of their electromagnetic and muonic content.

In order to separate the surviving hadronic background from the neutrino-induced showers, a Fisher discriminant analysis was performed using ROOT's Toolkit for Multivariate Data Analysis (TMVA)~\cite{TMVA}. The two variables $\Sem$ and $\Smu$ are combined into a single discriminant variable that maximizes the separation between the two populations.
Although the Fisher discriminant is optimal for Gaussian populations with equal covariance matrices, the shower observables considered here have asymmetric and relatively long-tailed distributions, so these assumptions are not strictly satisfied. We nevertheless use the Fisher discriminant because it is robust against overtraining with limited Monte Carlo statistics and provides a conservative baseline for comparison. More flexible and sophisticated multivariate methods, such as boosted decision trees or other neural network techniques, may improve separation in a future analysis with larger statistics, but are not required in this work to establish a proof of principle. 

\begin{figure}[htbp]
    \centering
    \includegraphics[width=0.49\textwidth]{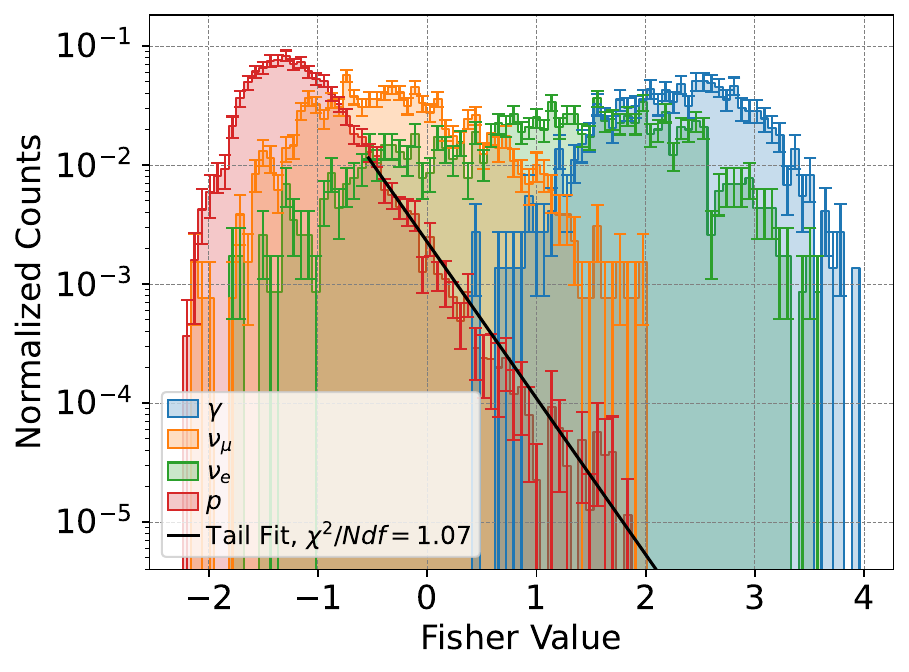}
    \includegraphics[width=0.49\textwidth]{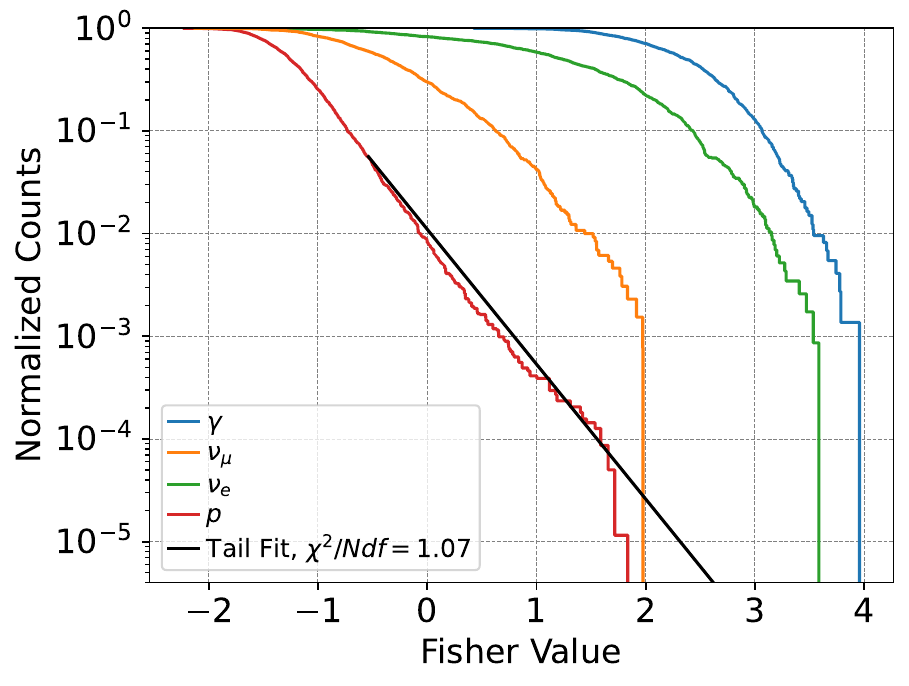}
    \caption{Fisher discriminant values derived from the variables $\Sem$ and $\Smu$ of electron neutrino (green), muon neutrino (orange), photon (blue), and proton (red) events with $E=10^{18}\,\mathrm{eV}$ and $\theta=30^\circ$. Left: Histogram of the Fisher values. An exponential curve (black solid line) has been fitted to the upper $10\%$ tail of the flux-weighted proton distribution. Right: Reverse cumulative histogram of the Fisher values.}
    \label{fig:Fisher_Values_Sem_Vs_Smu}
\end{figure}

The resulting Fisher variable distributions and reverse cumulative histograms for neutrino and proton events are shown in Figure~\ref{fig:Fisher_Values_Sem_Vs_Smu}. To estimate the background rejection achievable beyond the limits of the simulated statistics, an exponential fit is applied to the rightmost $10\%$ of the Fisher distribution for proton primaries, corresponding to the most neutrino-like events in the background sample.
The fit is shown in Fig.~\ref{fig:Fisher_Values_Sem_Vs_Smu}, and it is used to extrapolate the proton distribution towards higher Fisher values, allowing the selection of a Fisher threshold corresponding to a desired background rejection factor\footnote{We define the background rejection factor as the ratio of the number of background events before and after applying the Fisher threshold.}.

\subsubsection{Background-rejection for targeted searches}
\label{subsubsec:rejection_target}

%%%%%%%%%%%%%%%%%%%%%%%%%%%%%%%%

The background rejection required to reduce the expected number of background events to a sufficiently low level for a neutrino candidate search, depends on the search mode, with diffuse-flux searches generally requiring substantially stronger rejection than targeted multi-messenger searches.

In diffuse-flux analyses, rejection factors of the order of $10^{7}$--$10^{8}$ are typically required~\cite{AugerNuBkgRej_2013}. For an array covering $3000\,{\rm km^2}$, assuming the cosmic-ray flux measured by Auger~\cite{Auger_CR_Spectrum_ICRC2021}, this corresponds to fewer than one expected background cosmic-ray event every $5$--$50$~years in the energy and angular bin considered here\footnote{Such levels of rejection lie at the edge of or beyond what can be constrained with the present simulation statistics, and the diffuse channel is therefore treated primarily as a secondary benchmark in Section~\ref{sec:sensitivity}.}.

For targeted multi-messenger searches, the situation is different. An external alert allows for the definition of a restricted angular and temporal window, suppressing the cosmic-ray background and thereby relaxing the required background rejection. To set a quantitatively motivated target, we use as a reference benchmark the background level from the Earth-skimming (ES) neutrino searches of the Pierre Auger Observatory, for which the selection criteria are designed to yield fewer than one expected background event over approximately $50$ years of data taking~\cite{AugerNuBkgRej_2013}. Assuming conservatively one background event over this period, the probability of a background event occurring within a one-hour time window, representative of short-duration transients such as GRB-like flares and used here as a benchmark value, is

\begin{equation}
P^{\rm 1\,h}_{\rm CR} =
\frac{1}{50\times365\times24}
\simeq 2.3\times10^{-6}.
\label{eq:target_bkg}
\end{equation}
This probability is conservatively adopted as the benchmark target residual background, $n_{\rm bkg}=P^{\rm 1\,h}_{\rm CR}$, when searching for a vertical neutrino candidate associated with an astrophysical transient under such conditions.

The Pierre Auger Observatory surface detector records approximately $1200$ T3-triggered events per day~\cite{Auger_SD_Aperture_and_Trigger} which, over $N_{\rm yr}^{\rm ref}= 10$ years, yields $N_{\rm evt}^{[0^\circ, 60^\circ]}\simeq 2\times 10^6$ cosmic-ray events in the angular range $\theta\in[0^\circ, 60^\circ]$. This range corresponds to a solid angle 
\begin{equation*}
\Omega_{[0^\circ, 60^\circ]}=2\pi\,\left(1 - \cos(60^\circ)\right) = \pi\,{\rm sr}.
\end{equation*}
For a steady point-like source resolved with angular resolution $\Delta\alpha$, the corresponding solid angle is
$\Omega_{\Delta\alpha} \simeq 2\pi \, \left(1-\cos(\Delta\alpha)\right)$.
The expected number of background events within a search window $(\Delta\alpha, \,\Delta t)$ is then
\begin{equation}
N_{\rm bkg} = \frac{\Omega_{\Delta\alpha}}{\Omega_{[0^\circ, 60^\circ]}} \frac{\Delta t}{10\times365\times24} N_{\rm evt}^{[0^\circ, 60^\circ]} ,
\label{eq:Nbkg}
\end{equation}
which yields $ N_{\rm bkg}\simeq0.12$ for $\Delta t = 1$ hour and $\Delta \alpha = 3^\circ$, the latter being a deliberately conservative choice for the Auger SD angular resolution at the energies of interest~\cite{Auger_SD1500_AngRes}.

To reduce this background to the target residual of $n_{\rm bkg}=2.3\times 10^{-6}$, the required rejection factor is approximately
\begin{equation*}
r=\frac{N_{\rm bkg}}{n_{bkg}}\simeq 5\times 10^4
\end{equation*}
or, equivalently, a background contamination of $\varepsilon_{\rm bkg}\equiv 1/r = 2\times 10^{-5}$.
This value is adopted as the benchmark reference working point for all subsequent efficiency, effective-mass and effective-area results. For completeness, neighbouring contamination levels
($\varepsilon_{\rm bkg}=10^{-4}$, $10^{-5}$) are also reported, so that the sensitivity of the conclusions to the precise choice of search window can be assessed.

Two key aspects of this derivation should be noted. First, the one-hour window is representative of transients such as GRB-like flares. Longer-lived transients, spanning days to weeks, would require proportionally stricter rejection, while sub-hour alerts would relax it further. Second, the reference rate of $1200$ T3 events per day is an all-sky average and does not account for energy- or zenith-dependent variations in the trigger rate. These effects are not expected to change the required rejection factor significantly, and the resulting rejection factor $r=5\times 10^{4}$ therefore provides a conservative benchmark for this proof-of-principle study.

\subsubsection{Fisher analysis}
\label{subsubsec:fisher_analysis}

As explained above, the Fisher analysis is performed using simulated neutrino events within the region of interest shown in Fig.\,\ref{fig:Scatter_Sem_Vs_Smu}. The neutrino tagging efficiency is evaluated as a function of the slant depth between the interaction point and the ground, $DX$ in ${\rm g/cm^2}$, which strongly influences the shower properties at ground level and hence the discrimination power, with $DX$ defined as,
\[
DX = \frac{X_{\rm array}-X_{0}}{\cos\theta},
\]
where $X_0$ is the vertical atmospheric depth at which the neutrino interaction took place, $X_{\rm array}$ is the vertical atmospheric depth of the array ($875.5\,{\rm g/cm^2}$ in this work), and $\theta=30^\circ$. Larger values of $DX$ correspond to interactions occurring higher in the atmosphere.

\begin{figure}[htbp]
    \centering
    \includegraphics[width=0.55\textwidth]{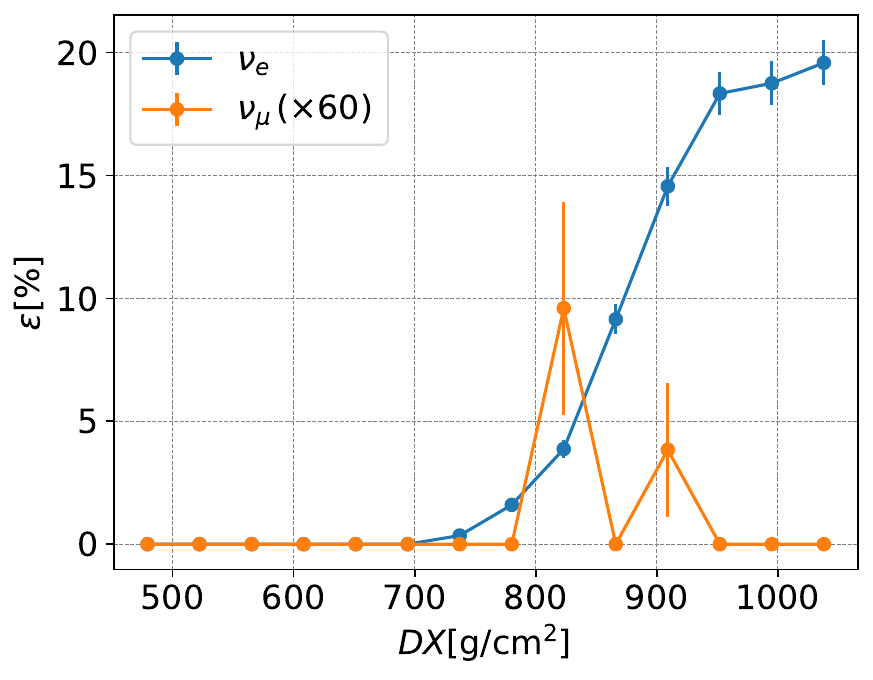}
    \caption{Neutrino identification efficiency (event trigger and tagging), $\varepsilon_\nu$, for $\nu_e$ (blue) and $\nu_\mu$ (orange) events with $E=10^{18}\,\mathrm{eV}$ and $\theta=30^\circ$ as a function of the slant depth $DX$ between the interaction point and the ground, for a background rejection factor $r=5\times10^{4}$ (see text) using the ideal discriminating variables $\Sem$ and $\Smu$. The $\nu_\mu$ efficiency has been multiplied by a factor of 60 for visibility.}
    \label{fig:Efficiency_vs_DX_Sem_Vs_Smu}
\end{figure}

An event is considered a successfully identified neutrino candidate if it passes the geometric cuts on the discriminant variables $\Sem$ and $\Smu$, obtained directly from the simulation in the ideal case, and its Fisher value exceeds the threshold for the desired background rejection factor. The resulting neutrino identification efficiency as a function of $DX$, for a rejection factor $r=5\times 10^4$, is shown in Fig.~\ref{fig:Efficiency_vs_DX_Sem_Vs_Smu} for both electron and muon neutrinos.

The triggering and tagging efficiency for electron neutrinos is null for very small values of $DX$, corresponding to interactions occurring very close to the ground, where the shower does not have sufficient atmospheric depth to develop and produce a significant electromagnetic signal at ground or even fulfil the trigger conditions. As $DX$ increases above $\sim750\,{\rm g/cm^2}$ the efficiency rises, and eventually approaches a plateau at around $900\,{\rm g/cm^2}$, with a maximum efficiency of approximately $20\%$ for a background rejection factor of $5\times10^4$. The plateau is attributed to showers that have reached maximum development before ground level, thus also achieving the maximum electromagnetic signal at the detector level. As $DX$ increases further reaching its limiting value of $\sim 1010\,{\rm g/cm^2}$ at $\theta=30^\circ$, the electromagnetic component starts to be attenuated before reaching the ground, leading to a decrease in the tagging efficiency. Interestingly, at the zenith angle $\theta=30^\circ$, the highest identification efficiency is achieved for neutrino interactions in the upper layers of the atmosphere. In this case, the smaller muonic component of $\nu_e$-induced showers provides the main source of discrimination against hadronic showers. For more inclined showers, however, the highest discrimination efficiency is typically obtained for neutrinos interacting deeper in the atmosphere \cite{Auger_EarthSkimming}.
As anticipated, the tagging efficiency for muon neutrinos is practically negligible across the full range of interaction depths. Consequently, only results for electron neutrinos are carried forward in the subsequent sections.

%%%%%%%%%%%%%%%%%%%%%%%%%%%%%%%%%%%%%%%%%%%%%%%%%%%%%%%%%%%%%%%%%%
\subsection{Experimental Implementation in Surface Detectors} 
\label{sec:exp_sd}

The pure shower component signals $\Sem$ and $\Smu$ discussed in the previous section are not directly measurable in surface detectors and therefore cannot be used as discriminating variables in a realistic experimental scenario. Instead, estimators for the electromagnetic and muonic components must be constructed from the signals recorded by surface detectors such as the Water-Cherenkov Detectors ($\SWCD$) and Surface Scintillator Detectors ($\SSSD$), as deployed in the AugerPrime upgrade~\cite{AugerPrime}.

The WCDs are sensitive to both the electromagnetic and muonic components of the shower, exhibiting a larger relative response to muons, primarily determined by the track length of the muon inside the water volume. In contrast, the signal generated by the electromagnetic component within WCDs depends on the number and energy distribution of secondary particles. Conversely, the SSDs are primarily sensitive to the electromagnetic component through scintillation light produced by charged particles with a relatively weak dependence on shower particle type. The combined use of WCD and SSD signals therefore provides the necessary indirect information regarding the electromagnetic-to-muonic ratio of the shower at ground level~\cite{Auger_WCD_SSD_Reconstruction}.

\begin{figure}[htbp]
    \centering
    \includegraphics[width=0.49\textwidth]{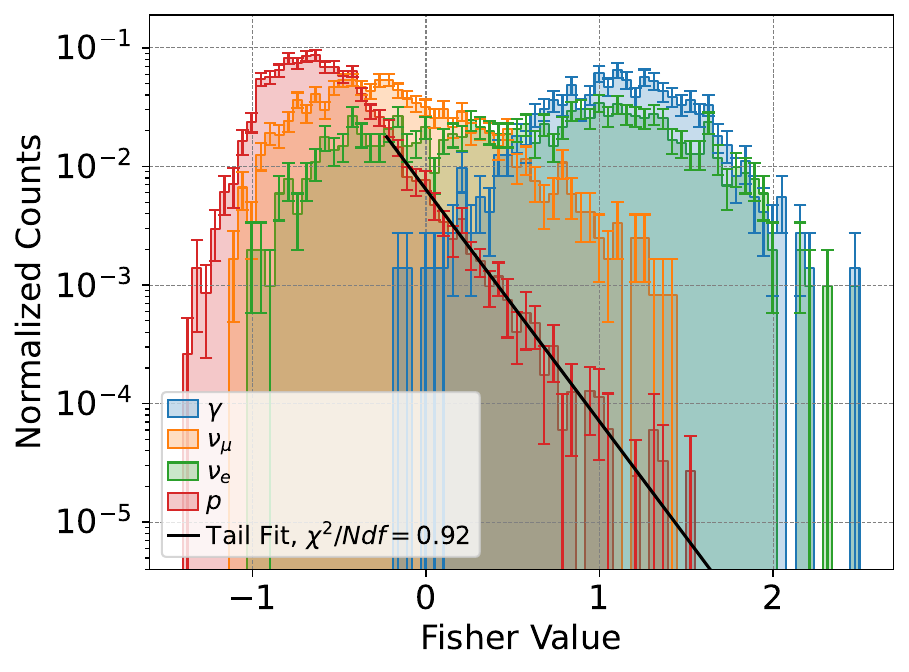}    
    \includegraphics[width=0.49\textwidth]{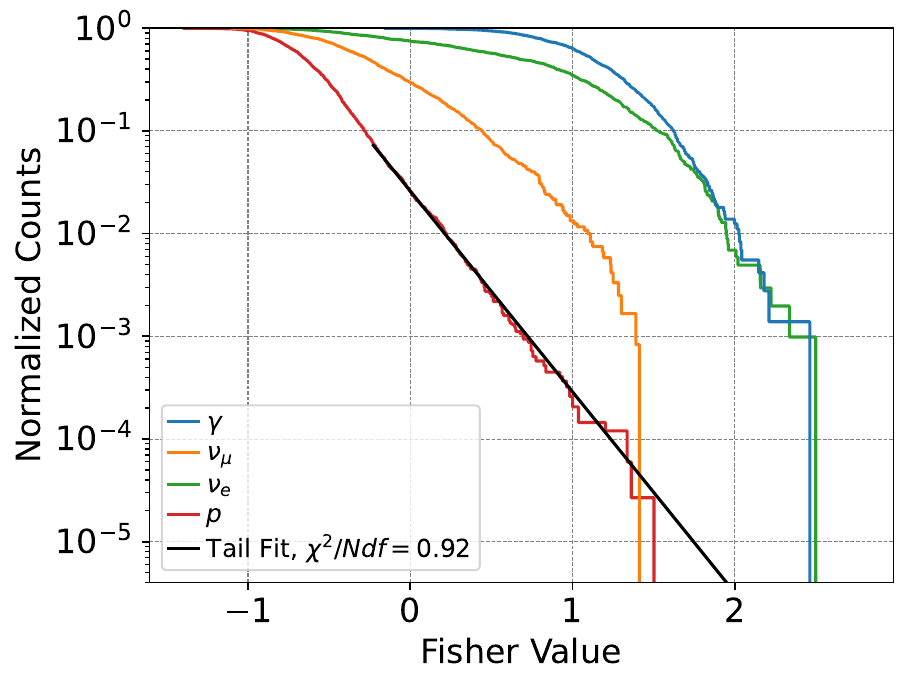}
    \caption{Fisher discriminant values derived from the observables $\SSSD$ and $\SWCD$ of electron neutrino (green), muon neutrino (orange), gamma (blue), and proton (red) events. Left: Histogram of the Fisher values. An exponential curve (black solid line) has been fitted to the upper $10\%$ tail of the flux-weighted proton distribution. Right: Reverse cumulative histogram of the Fisher values.}
    \label{fig:Fisher_Values_SSSD_Vs_SWCD}
\end{figure}

\begin{figure}[htbp]
    \centering
    \includegraphics[width=0.6\textwidth]{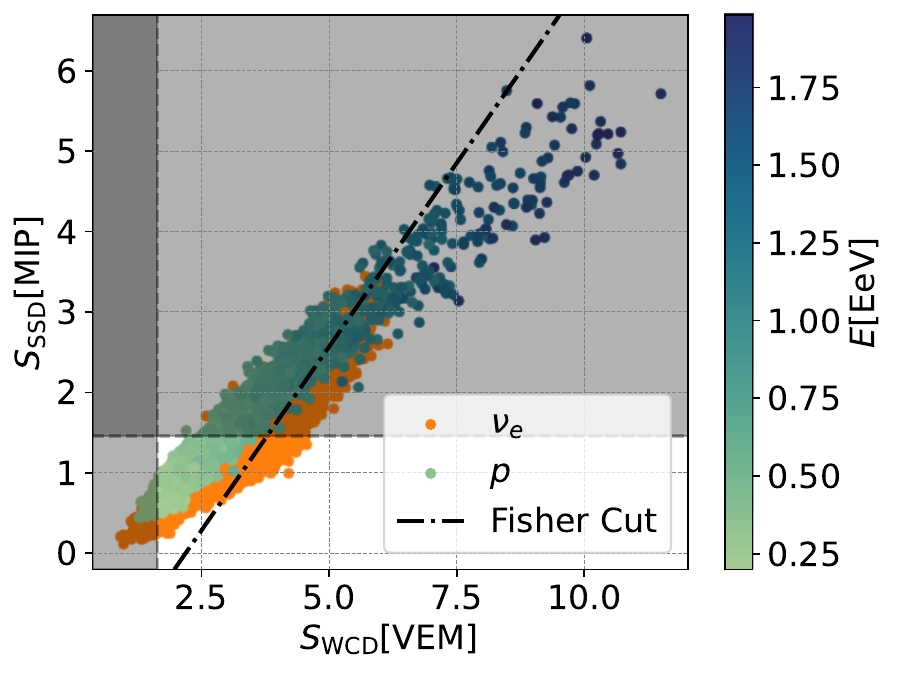}
    \caption{Scatter plot of the average values of $\SSSD$ and $\SWCD$ for stations located in a dense ring at $r=1000\,{\rm m}$. Electron neutrino events with $E=10^{18}\,\mathrm{eV}$ and $\theta=30^\circ$ are shown as orange markers, while proton events are color-coded according to their energy, with the color scale shown on the right. The dashed lines indicate the percentile cuts derived from the flux-weighted proton distributions, used to define the exclusion region for background events (shaded area). The dash-dotted line corresponds to the adjusted Fisher cut for a proton background rejection factor $r=5\times 10^{4}$.}
    \label{fig:SSD_vs_WCD_scatter}
\end{figure}

The Fisher discriminant analysis described in Section~\ref{subsec:ideal} was repeated using the average signals recorded by the WCDs, $\SWCD$, and SSDs, with $\SSSD$, extracted from the dense ring of virtual stations at $r=1000\,{\rm m}$ from the shower core.
The resulting Fisher variable distributions are shown in Fig.~\ref{fig:Fisher_Values_SSSD_Vs_SWCD}, and the corresponding scatter plot of $\SSSD$ versus $\SWCD$ in Fig.~\ref{fig:SSD_vs_WCD_scatter}. While there is more overlap between the two populations compared to the ideal case shown in Fig.\,\ref{fig:Scatter_Sem_Vs_Smu}, a region of interest can still be defined using the same criteria as in the ideal case.

\begin{figure}[htbp]
    \centering
    \includegraphics[width=0.55\textwidth]{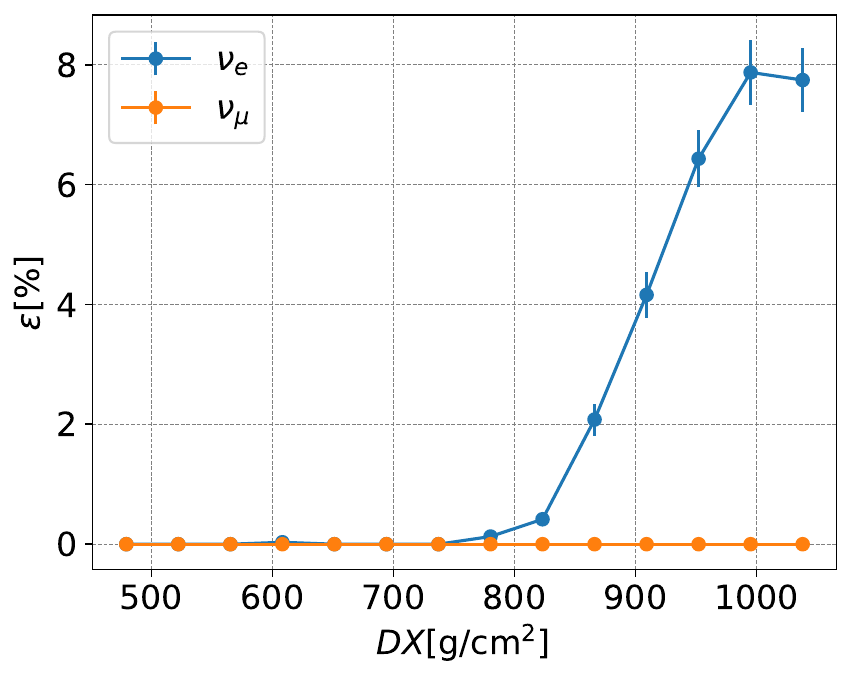}
    \caption{Neutrino identification efficiency (event trigger and tagging), $\varepsilon_\nu$, for $\nu_e$ (blue) and $\nu_\mu$ (orange) events with $E=10^{18}\,\mathrm{eV}$ and $\theta=30^\circ$ as a function of the slant depth $DX$ between the interaction point and the ground, for a background rejection factor $r=5\times10^{4}$ when using the discriminating variables $\SSSD$ and $\SWCD$.}
    \label{fig:Efficiency_vs_DX_SSSD_Vs_SWCD}
\end{figure}

The resulting neutrino identification efficiency (triggering and tagging) as a function of interaction slant depth is shown in Fig.~\ref{fig:Efficiency_vs_DX_SSSD_Vs_SWCD}. 
Compared to the ideal case, the maximum efficiency for electron-neutrino identification drops to $\sim8\%$, while preserving the trend of increasing efficiency with $DX$. This decrease in maximum achievable efficiency is expected, reflecting the intrinsic smearing introduced by detector responses and the limited ability to disentangle the electromagnetic and muonic components of the shower. Additionally, while in Section~\ref{subsec:ideal} the identification efficiency for muon neutrinos was significantly smaller than that for electron neutrinos, here the use of the experimentally accessible SSD and WCD observables leads to a complete suppression of muon-neutrino events after the selection cuts, reinforcing the conclusion that the vertical neutrino analysis performed in this work is exclusively sensitive to electron neutrinos.

With the identification efficiency $\varepsilon_\nu$ shown in Fig.\,\ref{fig:Efficiency_vs_DX_SSSD_Vs_SWCD}, the effective mass and effective area of the array as defined below can be calculated, allowing for an estimation of its sensitivity to UHE neutrinos, following the procedure described in Section~\ref{sec:sensitivity}. 

%%%%%%%%%%%%%%%%%%%%%%%%%%%%%%%%%%%%%%%%%%%%%%%%%%%%%%%%%%%%%%%%%%
\section{Sensitivity of a Ground Array to Neutrinos}
\label{sec:sensitivity}

Two complementary quantities can be obtained to translate the identification efficiency, $\varepsilon_\nu$, into an actual sensitivity to neutrinos. The effective mass, $M_{\rm eff}$, quantifies the amount of atmospheric target mass in which an interacting neutrino can produce a successfully identified shower, and is the relevant quantity for diffuse-flux estimates. The effective area, $A_{\rm eff}$, is the equivalent geometric collecting area of a perfectly efficient detector, and is the standard figure of merit for point-like and transient source searches~\cite{Auger_DownGoing}. Given the multi-messenger focus of this work, $A_{\rm eff}$ is treated as the primary result, while $M_{\rm eff}$ and the associated diffuse-flux benchmark are retained for completeness and for comparison with existing Auger limits.

\subsection{Effective mass}
\label{subsec:effmass}

For a fixed neutrino energy, $E_\nu$, the effective mass of the detector array, $M_{\rm eff}(E_\nu)$, is obtained by integrating the tagging efficiency over zenith angle $\theta$ and interaction slant depth $DX$:
\begin{equation}
    M_{\rm eff}(E_\nu)= 2\pi A \,\int_{\theta}\int_{DX}\sin \theta\,\cos\theta\,\, \varepsilon_\nu(E_\nu, \theta, DX)\, {\rm d}(DX) \,{\rm d\theta}\,,
    \label{eq:effMass}
\end{equation}
where $A$ is the geometrical area of the array. The factor $2\pi A \sin \theta \cos\theta$ accounts for the solid angle corresponding to neutrinos arriving with zenith angle $\theta$. The integration over $DX$ is performed using a cubic spline interpolation of the discrete simulation points. 

Although simulations were performed only at $\theta=30^\circ$, the 
discrimination efficiency is expected to improve with increasing slant depth, as suggested by the efficiency curves. For electron neutrinos (Figs.~\ref{fig:Efficiency_vs_DX_Sem_Vs_Smu} and \ref{fig:Efficiency_vs_DX_SSSD_Vs_SWCD}), the discrimination power increases as the interaction point moves farther from the ground, suggesting improved performance towards more inclined geometries within the vertical zenith-angle range. Conversely, the trigger efficiency is expected to decrease slightly towards $\theta=0^\circ$. To avoid relying on these competing and presently unquantified zenith-angle dependencies, we adopt a conservative approach and assume a constant overall efficiency over the narrow interval $\theta\in[30^\circ,\,35^\circ]$, where this approximation is expected to be reliable.

The resulting total effective mass values derived for both the ideal ($\Sem$, $\Smu$) and the experimentally accessible ($\SWCD$, $\SSSD$) sets of variables, for electron neutrinos, are reported in Table~\ref{tab:EffMass}. As expected they are smaller for the more realistic case of ($\SWCD$, $\SSSD$) compared to those obtained with the ideal set of variables ($\Sem$, $\Smu$).

\begin{table}[htbp]
\centering
\small
\renewcommand{\arraystretch}{1.2}
\setlength{\tabcolsep}{10pt}
    \begin{tabular}{ccc}
    \toprule
    $\varepsilon_{\mathrm{bkg}}$ &
    $M_{\mathrm{eff}}\,[\mathrm{g}]~(\Sem,\Smu)$ & 
    $M_{\mathrm{eff}}\,[\mathrm{g}]~(\SWCD,\SSSD)$\\
    \midrule
    $10^{-4}$ & $2.52 \times 10^{16}$ & $1.25 \times 10^{16}$\\
    $2\times10^{-5}$ & $1.74 \times 10^{16}$ & $6.79 \times 10^{15}$\\
    
    $10^{-5}$ & $1.31 \times 10^{16}$ & $4.53 \times 10^{15}$\\
    $6\times10^{-6}$ & $9.48 \times 10^{15}$ & $3.63 \times 10^{15}$\\
    \bottomrule
    \end{tabular}
\caption[Effective mass of the array for electron neutrinos]
{Effective mass of the array for electron neutrinos, integrated over the zenith angle range $\theta \in [30^\circ, 35^\circ]$ for several background contamination factors (first column), derived from the neutrino identification efficiency obtained using the ideal $\Sem$ and $\Smu$ discriminating variables (second column) and the $\SWCD$ and $\SSSD$ discriminating variables (third column).}
\label{tab:EffMass}
\end{table}

Once the effective mass is computed, it can be used to estimate the sensitivity of the array to a diffuse flux of neutrinos. The sensitivity is typically expressed in terms of the expected neutrino event rate, $\d N_\nu/\d t$, which can be calculated by convolving the effective mass with the incoming neutrino flux and the neutrino-nucleon cross section. It is given by:

\begin{eqnarray}
    \frac{\d N_\nu}{\d t} &=& \int^{E_{\nu,{\rm max}}}_{E_{\nu,{\rm min}}} 
    \frac{\d \Phi_\nu}{\d E_\nu}(E_\nu) \,
    \frac{\sigma_{\nu N,CC}(E_\nu)}{m_N} \,
    M_{\rm eff}(E_\nu) \,
    \d E_{\nu} \,,
    \label{eq:Sensitivity}
\end{eqnarray}
where $\d \Phi_\nu/\d E_\nu$ denotes the differential neutrino flux; $\sigma_{\nu N,CC}(E_\nu)$ the neutrino–nucleon cross section for the CC channel; $m_N$ is the proton mass; and $M_{\rm eff}(E_\nu)$ the effective mass\footnote{The effective mass was calculated including all neutrino--nucleon interactions, both charged-current and neutral-current processes. This leads to a slight underestimation of the effective mass, since neutral-current interactions, which account for roughly $30\%$ of all interactions at these energies, are not expected to contribute significantly to the neutrino detection efficiency. Their final states lack the energetic electromagnetic component that provides the main discrimination between cosmic-ray- and electron-neutrino-induced showers.} of the detector. We adopt the neutrino-nucleon charged-current cross section from~\cite{NuXSecs}. 
At the energy range considered in this analysis, the difference between neutrino and antineutrino cross sections is small, and therefore no explicit distinction is made between the two in the sensitivity calculation. Furthermore, the contribution from the Glashow resonance is not included in this work, since the energy range of interest is around $10^{18}\,$eV, significantly higher than the Glashow resonance energy of approximately $6.3 \times 10^{15}\,$eV.

Since no neutrino flux has yet been definitively measured at ultra-high energies, Equation~\ref{eq:Sensitivity} cannot be directly evaluated to provide a numerical value for the expected event rate. Instead, the sensitivity is expressed as an upper limit on the flux assuming zero observed events. Following standard statistical prescriptions for zero observed signal events and negligible expected background, the $90\%$ confidence level upper limit corresponds to $2.44$ expected events, a value derived from the Feldman-Cousins approach~\cite{FeldmanCousins1998}. To provide a baseline comparable to other observatory metrics, we express the scaling for a single expected event over an observation time $\Delta t$ as: 
\begin{equation}
    E_\nu^2\,\frac{\d \Phi_\nu}{\d E_\nu} = 
    E_\nu^2\,\left[ 
    \frac{\sigma_{\nu N,CC}(E_\nu)}{m_N}\,
    M_{\rm eff}(E_\nu) \,
    \Delta E_\nu \, \Delta t
    \right]^{-1}.
    \label{eq:NuFlux}
\end{equation}
An estimation of the limit on the diffuse flux of neutrinos can be obtained, taking into account an observation time of $\Delta t = 16$ years, comparable to the first data-taking phase of the Pierre Auger Observatory, and an energy bin of width $\Delta \log_{10} E = 0.5$ centred at $10^{18}\,{\rm eV}$. At ultra-high energies, the interaction physics and subsequent shower development are essentially identical for neutrinos and antineutrinos. In charged-current interactions, an electron neutrino produces an energetic electron that initiates an electromagnetic shower, while an electron antineutrino produces a positron that generates a nearly identical cascade. For this reason, the sensitivity is computed for the combined $\nu_e + \bar{\nu}_e$ flux, corresponding to a factor of two increase in the expected event rate. 

The most stringent background contamination considered here, $\varepsilon_{\rm bkg}=6\times10^{-6}$, comparable to that employed by the Pierre Auger Observatory in Earth-skimming neutrino searches~\cite{Auger_Neutrino_2024}, yields an upper limit on the diffuse neutrino flux of
\[
E_{\nu}^2\,\frac{\d \Phi_\nu}{\d E_\nu}
\lesssim 2.03\times10^{-7}\,
{\rm GeV\,cm^{-2}\,s^{-1}\,sr^{-1}},
\]
when restricting the analysis to the $30^\circ$--$35^\circ$ zenith-angle interval and using the realistic variable set, $(\SWCD,\SSSD)$. This limit is approximately a factor of fifteen weaker than that reported by the Pierre Auger Observatory,
$E_\nu^2\,{\d \Phi_\nu}/{\d E_\nu}\lesssim
1.3\times10^{-8}\,{\rm GeV\,cm^{-2}\,s^{-1}\,sr^{-1}}$,
in an energy bin $\Delta\log_{10}E=0.5$ centred at $10^{18}\,{\rm eV}$, where the sensitivity is dominated by the ES channel~\cite{Auger_Neutrino_2024}. 
Thus, when considering limits to diffuse fluxes and zenith-angle intervals of comparable width, the proposed vertical method does not achieve a sensitivity competitive with the Auger Earth-skimming approach, reinforcing its primary relevance for point-source and transient searches.

The larger angular coverage potentially accessible to the vertical channel can, however, substantially increase its integrated sensitivity. Extrapolating the result to the full $25^\circ$--$60^\circ$ zenith-angle interval, under the assumption that the identification efficiency remains constant with zenith angle, yields
\[
E_{\nu}^2\,\frac{\d \Phi_\nu}{\d E_\nu}
\lesssim 2.81\times10^{-8}\,
{\rm GeV\,cm^{-2}\,s^{-1}\,sr^{-1}}.
\]
This extrapolated value is only a factor of two weaker than the Auger limit at $10^{18}\,{\rm eV}$. This result indicates that, although the method is not competitive with the Earth-skimming channel over a narrow zenith-angle interval, the substantially larger solid angle potentially available to the vertical search could compensate for its lower sensitivity per unit angular coverage. Moreover, the observed increase in neutrino detection efficiency for interactions occurring higher in the atmosphere, i.e. at larger $DX$, suggests that the performance may improve towards larger zenith angles. A dedicated simulation across the full vertical zenith-angle range is therefore required to quantify this dependence and establish whether the projected sensitivity can be realised.

%%%%%%%%%%%%%%%%%%%%%%

\subsection{Effective area for point-like sources}
\label{subsec:effarea}

An alternative and, for the present purposes, more informative means to assess the sensitivity of the array to neutrinos is the effective area, $A_{\rm eff}(E_\nu, \theta)$, given by 
\begin{equation}
    A_{\rm eff}(E_\nu, \theta) = \int_{DX}\int_{A}  \varepsilon_\nu(E_\nu, \theta, DX) \,\cos\theta\, \frac{ \sigma_{\nu N,CC}(E_\nu)}{m_N}  \,\d (DX)\, \d A,
    \label{eq:effArea}
\end{equation}
Because $A_{\rm eff}$ relates linearly to the expected number of events from a flux from a point-like source, it is the standard quantity used to compare directional neutrino searches across experiments~\cite{Auger_DownGoing}. The values obtained in this work at $E_\nu=10^{18}\,{\rm eV}$ are reported in Table~\ref{tab:EffArea}. At the reference $\varepsilon_{\rm bkg}=2\times 10^{-5}$ the $(\SSSD,\SWCD)$ analysis yields $A_{\rm eff}\simeq 1.6\times 10^{-2}\,{\rm km^2}$, while the idealised $(\Sem,\Smu)$ analysis reaches a larger effective area $A_{\rm eff}\simeq 4.0\times 10^{-2}\,{\rm km^2}$.

\begin{table}[htbp]
    \centering
    \small
    \setlength{\tabcolsep}{10pt}
    \begin{tabular}{ccc}

    \toprule &
    $A_{\rm eff}(E_\nu, \theta),[\mathrm{km^2}]$ & $A_{\rm eff}(E_\nu, \theta),[\mathrm{km^2}]$ \\
    $\varepsilon_{\mathrm{bkg}}$ & $(\Sem, \Smu)$ & $(\SWCD, \SSSD)$\\
    \midrule
    $10^{-4}$ & $ 5.8 \times 10^{-2}$ & $2.9 \times 10^{-2}$\\
    $2\times10^{-5}$ & $ 4.0 \times 10^{-2}$ & $1.6 \times 10^{-2}$\\ 
    $10^{-5}$ & $ 2.6 \times 10^{-2}$ & $7.4 \times 10^{-3}$ \\
    
    \bottomrule
    \end{tabular}

    \caption[Electron neutrino effective area obtained from the present method for background rejection factors of $10^{4}$ and $10^{5}$]
    {Electron neutrino effective area at \(E_\nu=10^{18}\) eV and \(\theta=30^\circ\) obtained from the present method using the $\Sem$ and $\Smu$, and the $\SWCD$ and $\SSSD$ variables for several background rejection factors.}
    \label{tab:EffArea}
\end{table}

\begin{figure}[htbp]
    \centering
    \centering
    \includegraphics[width=0.69\textwidth]{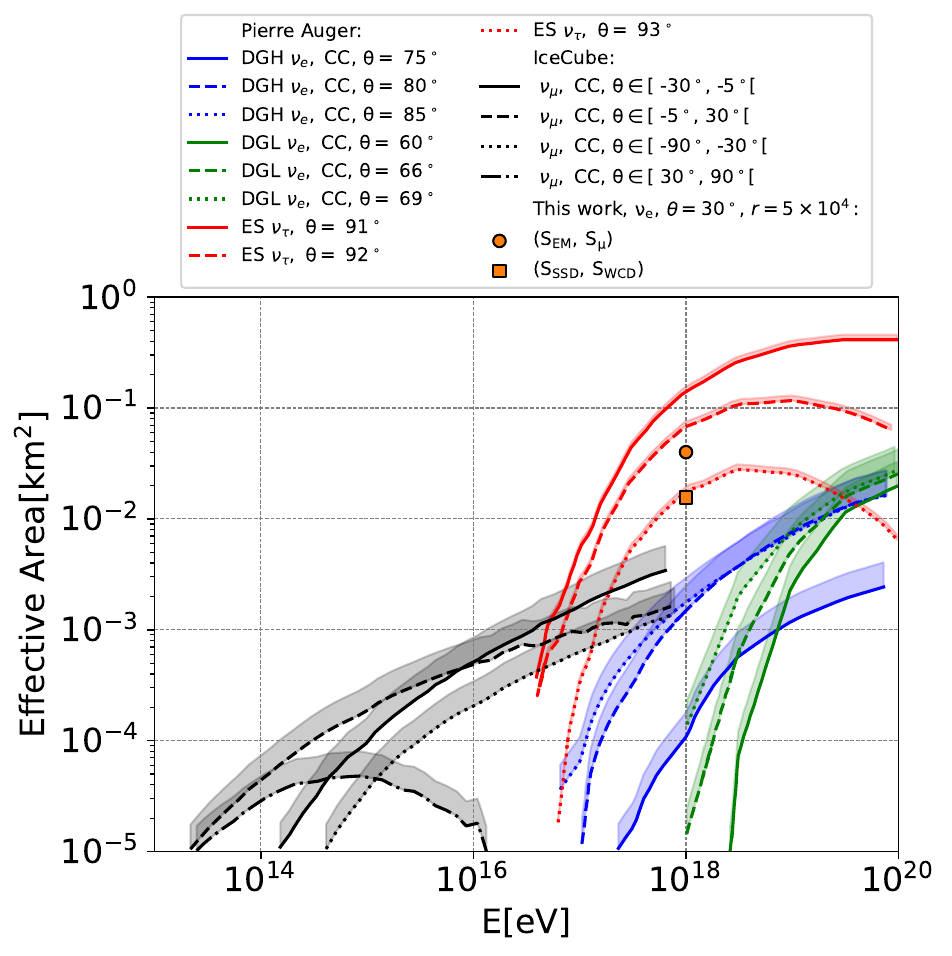}
    \caption[Comparison of the neutrino effective area with current results, as a function of energy]
    {Total effective area as a function of neutrino energy for the vertical electron-neutrino search using $(\Sem,\Smu)$ and $(\SSSD,\SWCD)$ at $\varepsilon_{\rm bkg}=2\times 10^{-5}$, compared with Pierre Auger results for DGL and DGH electron neutrinos and ES tau neutrinos, and with IceCube results for muon-neutrino charged-current interactions~\cite{Auger_DownGoing}. Shaded bands indicate the estimate of the maximum effective area achieved by the current analyses of Auger and IceCube by saturating their neutrino identification to $100\%$ efficiency (see text).
    }
    \label{fig:EffectiveArea_Energy_Comparison}
\end{figure}

\begin{figure}[htbp]
    \centering
    \centering
    \includegraphics[width=0.69\textwidth]{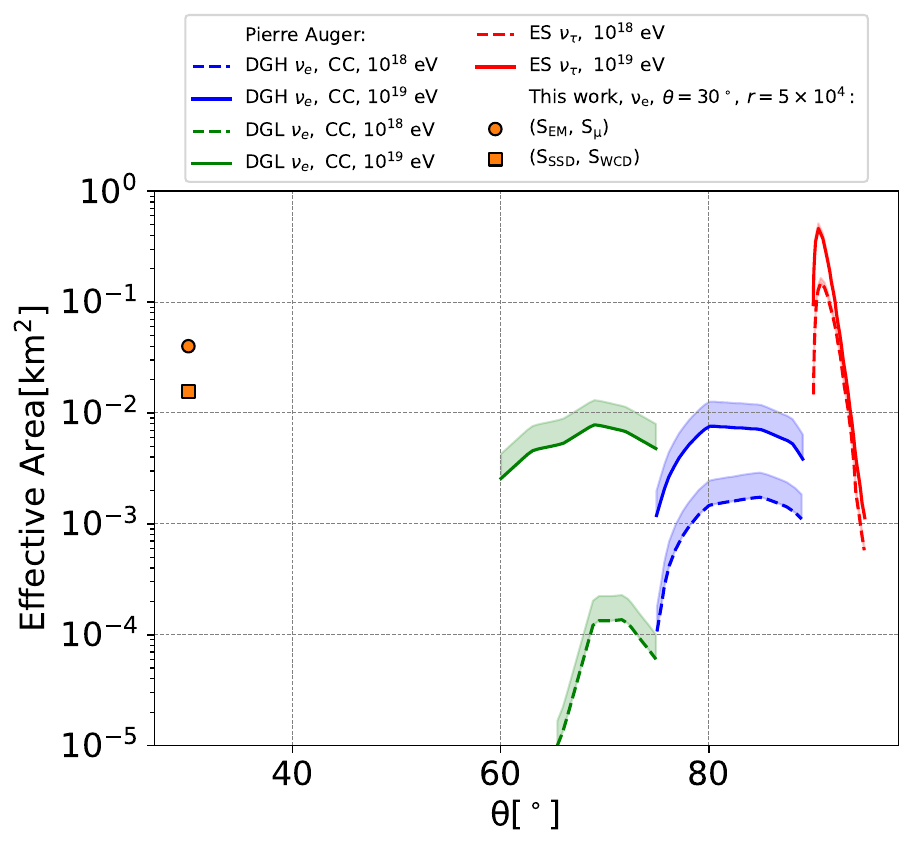}
    \caption[Comparison of the neutrino effective area with current results, as a function of zenith angle]
    {Total effective area as a function of zenith angle for the vertical search using $(\Sem,\Smu)$ and $(\SSSD,\SWCD)$ at $\varepsilon_{\rm bkg}=2\times 10^{-5}$, compared with Pierre Auger results for the DGL, DGH and ES channels~\cite{Auger_DownGoing}. Shaded bands as in Figure~\ref{fig:EffectiveArea_Energy_Comparison}.}
    \label{fig:EffectiveArea_Theta_Comparison}
\end{figure}

Figures~\ref{fig:EffectiveArea_Energy_Comparison} and~\ref{fig:EffectiveArea_Theta_Comparison} compare these results with the published Auger effective areas for the downward-going low (DGL, $\theta\in[60^\circ,\, 75^\circ]$), downward-going high (DGH, $\theta\in[75^\circ,\,90^\circ]$) and Earth-skimming (ES, $\theta\in[90^\circ,\,95^\circ]$) channels, and with the IceCube effective area for muon-neutrino charged-current interactions~\cite{Auger_DownGoing}. All results obtained with the present method are shown at $\varepsilon_{\rm bkg}=2\times10^{-5}$ and without integration over zenith angle; instead, the effective area is evaluated directly at $\theta=30^\circ$ to ensure a consistent comparison. The shaded bands indicate a rescaling of the published Auger and IceCube effective areas to a common $100\%$ identification efficiency, assuming an efficiency for the Earth-Skimming channel of $90\%$ and a representative $60\%$ efficiency for the remaining cases \cite{Auger_EarthSkimming}. The rescaling accounts for the larger and hence more stringent rejection factors used in the Auger and IceCube analyses, which were not optimized for the lower rejection levels required for short-duration, point-like sources. It also enables a more direct comparison of the intrinsic effective areas by mitigating differences in identification efficiencies arising from the analysis-specific cuts.

Two observations follow from this comparison. First, at $\varepsilon_{\rm bkg}=2\times 10^{-5}$, the rejection factor appropriate to a one-hour point-source follow-up, the vertical channel yields an effective area of a comparable order of magnitude to those of the established Auger ES, DGH, and DGL channels in the overlapping energy range, and is complementary in zenith angle as it populates the previously unexploited $\theta<60^\circ$ angular range of the field of view. Second, care should be taken not to over-interpret this comparison. The reference Auger and IceCube curves are derived from full instrument response functions spanning broad energy and zenith ranges, with more stringent trigger conditions for the case of Auger especially for the DGH and DGL channels \cite{Auger_EarthSkimming}, whereas the present result is a single-energy, single-zenith proof of principle with a simplified and more relaxed trigger conditions and idealised signal extraction. The appropriate conclusion is therefore not that the vertical channel supersedes or outperforms existing searches, but that it has the potential to provide a comparable instantaneous sensitivity for electron neutrinos in a multi-messenger alert scenario, at no additional instrumental cost in the case of Auger, beyond the AugerPrime SSD layer already deployed. The strength of this approach lies in the strong reduction of the cosmic-ray background provided by the temporal and directional coincidence with a multi-messenger alert. This background suppression can compensate, to some extent, for the intrinsically lower discrimination power of the techniques employed, making the identification of neutrino-induced events possible despite their less distinctive signatures. Additionally, the geometrically projected area entering in Eq.\,\eqref{eq:effArea} is larger by a sizeable factor of $\cos(30^\circ)/ \cos \theta$ for the vertical case compared to the inclined channels. 
Finally, the trigger efficiency for \textit{vertical} neutrino-induced showers is also expected to be higher than for \textit{inclined} events\footnote{This expectation is supported by the lower energy threshold at which the Auger SD reaches full trigger efficiency for vertical showers~\cite{PierreAuger:2014jss}.}, thereby further increasing the effective area of the vertical channel.

%%%%%%%%%%%%%%%%%%%%%%%%%%%%%%%%%%%%%%%%%%%%%%%%%%%%%%%%%%%%%%%%%%%
\section{Photon-induced showers as a background to neutrino identification}
\label{sec:neutrals}

Electron neutrino-induced and gamma-ray-induced air showers share similar characteristics, primarily due to their dominant electromagnetic components. In both cases, the shower development is driven by electromagnetic cascades, resulting in particle distributions rich in electrons, positrons, and photons. Consequently, these showers exhibit similar lateral distributions and temporal profiles at ground level, and produce comparable signals in surface detectors. This makes it difficult to distinguish between neutrino-induced and photon-induced showers using surface detector information alone. This similarity is particularly strong for charged-current electron neutrino interactions, where the outgoing electron initiates an electromagnetic cascade that closely resembles that of a gamma-ray-induced shower.

For vertical showers, the relatively small atmospheric thickness further reduces the differences in shower development between neutrino-induced and photon-induced events. This similarity is also apparent in Figs.~\ref{fig:Fisher_Values_Sem_Vs_Smu} and~\ref{fig:Fisher_Values_SSSD_Vs_SWCD}, where photon-induced showers, shown in blue, populate a region similar to that of neutrino-induced showers. The method introduced in this work should therefore be interpreted more generally as a technique to identify vertical air showers initiated by neutral particles, rather than as a method to uniquely identify neutrinos. In the context of multi-messenger observations, this is not necessarily a limitation, as both photons and neutrinos are neutral messengers that point back to their astrophysical sources. Additional strategies would be required to distinguish between the different types of neutral primaries responsible for the observed showers, which is beyond the scope of this study.

%%%%%%%%%%%%%%%%%%%%%%%%%%%%%%%%
\section{Conclusions}
\label{sec:conclusions}

In this work, we have investigated the feasibility of extending searches for UHE neutrinos with surface detector arrays to the vertical zenith-angle range, $\theta<60^\circ$, which is not exploited by existing neutrino searches. The method relies on the enhanced electromagnetic component of showers initiated by charged-current electron-neutrino interactions, which provides discrimination against the hadronic cosmic-ray background. We have demonstrated this separation both using the idealised electromagnetic and muonic shower components, $(S_{\rm em},S_\mu)$, and with experimentally accessible observables based on the complementary response of scintillator and water-Cherenkov detectors, $(S_{\rm SSD},S_{\rm WCD})$.

The discrimination at $\theta=30^\circ$ is found to be driven almost exclusively by electron neutrinos and depends strongly on the atmospheric depth available for shower development and hence on zenith angle. For the reference background contamination $\varepsilon_{\rm bkg}=2\times10^{-5}$, motivated by a one-hour multi-messenger follow-up within a $3^\circ$ angular window, the realistic SSD--WCD analysis yields an electron-neutrino effective area of $A_{\rm eff}\simeq1.6\times10^{-2}\,{\rm km^2}$ at $10^{18}\,{\rm eV}$ and $\theta=30^\circ$. The corresponding idealised analysis reaches $A_{\rm eff}\simeq4.0\times10^{-2}\,{\rm km^2}$. These values are of the same order of magnitude as the effective areas obtained for inclined-shower neutrino searches in the overlapping energy range, while extending the field of view to the complementary $\theta<60^\circ$. The comparison should be interpreted in the context of the present proof-of-principle study, which focuses on a single energy and zenith angle and provides an initial assessment rather than a comprehensive experimental analysis.

The reduced background associated with the temporal and directional coincidence of a multi-messenger alert is central to the viability of the \textit{vertical channel}. For diffuse searches, where substantially stronger background suppression is required, the method is less competitive over zenith-angle intervals of comparable width. At $\varepsilon_{\rm bkg}=6\times10^{-6}$, the realistic analysis restricted to $\theta\in[30^\circ,35^\circ]$ yields a diffuse-flux sensitivity approximately a factor of $15$ weaker than the current Pierre Auger limit around $10^{18}\,{\rm eV}$. The larger solid angle potentially accessible to the vertical channel could partly compensate for this difference: extrapolating the identification efficiency over $\theta\in[25^\circ,60^\circ]$ results in a sensitivity only a factor of two lower than the Auger limit. This estimate relies, however, on the unverified assumption of a zenith-independent efficiency and should therefore be regarded only as an indication of the potential gain from the larger angular coverage.

Several ingredients will be required to turn this proof of principle into a realistic experimental search. The present simulations are restricted to $E_\nu=10^{18}\,{\rm eV}$ and $\theta=30^\circ$, and substantially larger simulation samples are required to determine the energy and zenith-angle dependence of the identification efficiency. The trigger implementation is also simplified with respect to the full Auger SD trigger and uses an effective station threshold calibrated to reproduce its reported efficiency for benchmark proton showers. A complete analysis should incorporate the full trigger and detector response, realistic reconstruction of the ground signals, and more powerful multivariate classifiers, whose performance can be robustly assessed once sufficient simulation statistics are available.

Finally, the surface-detector observables considered here do not provide an efficient means of separating electron-neutrino-induced from photon-induced showers, owing to their similarly large electromagnetic components. The proposed selection should therefore be regarded more generally as identifying neutral-particle (non-hadronic) candidates. This broader sensitivity makes the method even more relevant to multi-messenger searches, where temporal and directional information from an external alert can complement the surface-detector observables and help constrain the astrophysical origin of a candidate event.

Overall, these results establish the feasibility of using vertical air showers as a complementary UHE neutral-particle detection channel in surface arrays. For an AugerPrime-like experiment, the existing combination of WCD and SSD measurements provides the information required to exploit this channel, without additional detector hardware. Extending the analysis across energy and zenith angle will determine the extent to which the presently unused vertical field of view can enhance the multi-messenger reach of current and future surface detector arrays.

\acknowledgments
We would like to thank the Pierre Auger Collaboration for the valuable discussions during the development of this work, and in particular Carola Dobrigkeit for her careful reading of the manuscript.
This work has been funded by Fundação para a Ciência e Tecnologia, Portugal, under project \url{https://doi.org/10.54499/2024.06879.CERN}.
P.~C. is grateful for the financial support by FCT under UI/BD/153576/2022.
This work has also received financial support from:
Ministerio de Ciencia, Innovaci\'on y Universidades/Agencia Estatal de Investigaci\'on, MICIU/AEI /10.13039/501100011033, Spain
(PID2022-140510NB-I00, PCI2023-145952-2, RYC2019-027017-I, CNS2024-154676, and Mar\'\i a de Maeztu grant CEX2023-001318-M); Xunta de Galicia, Spain (CIGUS Network of Research Centers, and Consolidaci\'on ED431C-2025/11); and the European Union (Galicia Feder 2021-2027 Program and Horizon Europe Research and Innovation Program under Grant Agreement No. 101131928 - ACME).

\clearpage 
\appendix

\bibliographystyle{JHEP}
\bibliography{references.bib}

\end{document}